\documentstyle[prbbib,preprint,tighten,aps]{revtex}
\begin{document}
\title
{\bf Resonant and coherent transport through Aharonov-Bohm
interferometers with coupled quantum dots }

\author{V. Moldoveanu$^{1,2}$, M. \c{T}olea$^1$, A. Aldea$^1$,
and B. Tanatar$^3$}
\address{$^1$National Institute of Materials Physics, P.O. Box MG-7,
Bucharest-Magurele, Romania\\
$^2$ Centre de Physique
  Th\'{e}orique-CNRS, Case 907 Luminy,
                               13288 Marseille Cedex 9, France\\
$^3$Department of Physics, Bilkent University, Bilkent, 06800
Ankara, Turkey}

\maketitle

\begin{abstract}
A detailed description of the tunneling processes within Aharonov-Bohm 
(AB) rings containing two-dimensional quantum dots is presented. 
We show that the electronic propagation through the interferometer is 
controlled by the spectral properties of the embedded dots and by 
their coupling with the ring.
The transmittance of the interferometer is computed by 
the Landauer-B\"uttiker formula.   
Numerical results are presented for an AB interferometer containing 
 two coupled dots. The charging diagrams for a double-dot
interferometer and the Aharonov Bohm oscillations 
are obtained, in agreement with the recent experimental results of 
Holleitner {\it et al}. [Phys. Rev. Lett. {\bf 87}, 256802 (2001)] 
We identify conditions in which the system shows Fano line shapes.
The direction of the asymetric tail depends on the capacitive coupling and on the 
magnetic field. We discuss our results in connection 
with the experiments of Kobayashi {\it et al} [Phys. Rev. Lett. {\bf 88}, 256806 (2002)]
 in the case of a single dot.  

%The coherence 
%properties of various geometries are studied. 

\end{abstract}
\maketitle
%\newpage
%\par
\section{Introduction}

The electronic transport through Aharonov-Bohm rings with embedded
quantum dots (QD's) is a new subject in mesoscopic  physics
whose complexity competes with the already 'classical' problem of 
persistent currents in closed loops.

Inserting one dot in a ring Yacoby {\it et al}. \cite{1} studied 
for the first time the transport properties of such systems. The 
observed Aharonov-Bohm (AB) oscillations of the source-drain 
signal as the magnetic field is varied proved that the tunneling 
current through the dot is partially coherent.  
The experiment presented a striking behavior of the
transmittance phase as a function of the gate voltage applied on the 
dot: at each transmittance peak the phase jumps by $\pi$.
Due to the two-lead geometry used in this experiment the conductance 
obeys the Onsager relations and as shown in \cite{LB}, this imposes 
a rigidity of the transmittance phase (0 or $\pi$). 
Later on Shuster {\it et al}. \cite{2,3} employed a many lead geometry, 
the phase evolution being obtained
as well as the expected Aharonov-Bohm oscillations. 
The experimental geometry was generalized by Holleitner 
{\it et al}. \cite{4} who measured the current through a double dot 
AB interferometer (one QD in each arm of the ring). 
The main achievement of their setup is that the dots can be coherently
coupled and hence the transport becomes more complex. 
They have also found AB oscillations of the current and emphasized the 
formation of coherent molecular states in the two dots.
Finally, a recent experiment \cite{4fire} with two-dots AB
ring was performed in a four-lead geometry, the measured transmittance 
showing peaks in several regimes of the capacitive 
coupling of the ring.  Notably, the phase of the transmittance 
presents the same increment with $\pi$ when one dot is set to resonance 
and the capacitive coupling of the second dot is varied around a peak.

A closely related problem is the Fano feature of AB interferometers.
As reported in Ref.\cite{F1} a one dot AB interferometer shows 
asymmetric line shapes for the transmittance as a function of 
the plunger gate voltage, the typical proof of the Fano 
effect \cite{Fano,Nockel,Wulf}, namely the interference between states 
belonging to continuous and discrete spectra. 

The transport properties of AB interferometers containing QD's were 
theoretically studied  by two techniques, the 
scattering approach and the Keldysh formalism in the tight binding 
picture. The scattering theory was successfully used 
in \cite{HW1,HW2,Ha} to describe the physics of one-dot 
interferometers, including specific properties of the
transmittance phase. The $S$ matrix of the scattering problem 
is computed by writing the Born expansion for the $T-$operator, 
the conductance being thereafter obtained via the 
Landauer-B\"uttiker formula.

In the tunneling picture the net current from one lead to another is 
computed by perturbation theory and non-equilibrium Green 
function techniques. Within this approach one can discuss in detail 
the co-tunneling spin-dependent processes and finite bias 
transport \cite{YK}.
As discussed recently in Ref.\cite{YK1} both approaches are equivalent,
in spite of the differences between the Hamiltonians (in the tunneling 
picture the coupling between the ring and dots does not appear 
explicitly).

The way in which the experiments with AB interferometers can 
indeed provide the transmittance phase is a subtle point that 
involve the explicit geometry of the leads used to break the 
unitarity of the two-lead system (see \cite{W1,A1,A2,TJP}).

In the present work we study systematically the tunneling and 
coherence properties of AB interferometers with QD's, particular 
attention being payed to the geometry used in Ref. \cite{4}. The 
idea behind the calculations presented below is the following. 
The transmittance of the interferometer as a whole is first 
related to its Green function, by the Landauer-B\"uttiker formula. 
Secondly, it is shown that this Green function can be expressed in 
terms of two Green functions that describe {\it separately} 
the ring and the dots system in the presence of the leads. 
The lead-ring, lead-dots and ring-dot couplings 
appear as non-hermitian self-energies of an effective Hamiltonian.
The latter is obtained by the Feschbach formula \cite{Fes,RMP} 
which is a useful tool when dealing with Hamiltonians
of coupled subsystems. We point out that this step is necessary
in order to obtain detailed information about the complex processes
within the interferometer.
The resonant transport through the device is discussed in 
connection with the spectral properties of the dots system embedded 
in the interferometer.
Our approach shows clearly that the important 
role in the {\it resonant} transport processes is played by the 
dots inserted in the ring, the latter providing in turn the 
suitable geometry for quantum {\it coherence}. 

We do not consider in this paper the Coulomb repulsion 
because interaction effects on the transport properties of single and
coupled dots were studied extensively in the previous 
papers \cite{PRB1,PRB2,PE} and all the analysis presented there 
remains valid here.
The Coulomb interaction can be however easily included in our formalism
in the Hartree approximation and the charging effects are 
satisfactorily described by this approach
(see \cite{HW1} for a similar discussion of the interaction effects
in a one-particle approximation).
The main topics we consider in this work are the tunneling and
coherence properties of AB interferometers. The Kondo-type effects 
which are a subject in itself are not discussed here.

The formalism is presented in Section II. Numerical results are 
discussed in Section III in connection with the experimental findings,
a qualitative agreement being found. 
Since we have considered two-dimensional quantum dots 
%accommodating 
%up to 20 electrons as in the experimental setup of Holleitner {\it et
%al}. \cite{4} 
the magnetic field dependence
of the eigenvalues of the coupled dots system is no longer 
negligible as in the case of a dot modeled by a single site. 
It will turn out that the drift of the levels in magnetic field affects the
interferometer transport properties.
Moreover, the interferometer regime of the device (namely the one 
that exhibits AB oscillations) is more difficult to reveal. 
Section IV summarize the main results and ends the paper. 

\section{Formalism}

This section contains the theoretical framework we use to study 
the electronic transport in Aharonov-Bohm interferometers with 
coupled quantum dots. The Hamiltonians are written in 
the tight-binding (TB) representation which is particularly useful both
for describing complex geometries and performing numerical computations.
We consider a general interferometer that consists of an arbitrary 
number of two-dimensional (coupled) quantum dots embedded in a 
1D mesoscopic ring having $N$ sites. 
Some of these sites are shared with the dots, which are  
coupled to each other by tunneling Hamiltonians simulating the
tunable barriers patterned in experiment. The quantum dots  
are described as finite two-dimensional (2D) plaquettes.

The electrons reach and leave the interferometer 
through ideal one-channel semi-infinite leads attached  
on the ring or directly on the dots.
The Hamiltonian of the whole system has the form
\begin{equation}\label{H}
H=H^I+H^L+H^{LI}_{{\rm tun}},
\end{equation}
with 
\begin{equation}\label{HI}
H^I=H^D+H^R+H^{RD}_{{\rm tun}}.
\end{equation}
$H^I$ is the Hamiltonian of the interferometer,
$H^L$ and $H^R$ describe the leads and the truncated ring, 
i.e. what is left from it after removing the dots (the notations can be identified as well 
from Fig.1 which represents a double dot interferometer). 
The magnetic flux through the ring appears
in $H^R$ in the Peierls representation as magnetic phases attached to the hopping constants 
along the 
truncated ring. Their explicit form is obtained by using for example the Landau gauge. 
$H^{LI}_{{\rm tun}}$ and $H^{RD}_{{\rm tun}}$ are the
lead-interferometer and ring-dots tunneling Hamiltonians:
\begin{eqnarray}
H^{LI}_{{\rm tun}}&=&H^{LI}+H^{IL}=
\tau_L\sum_{\alpha}(|0_{\alpha}\rangle\langle \alpha|+
|\alpha\rangle\langle 0_{\alpha}| ),\\
H^{RD}_{{\rm tun}}&=&H^{DR}+H^{RD}
=\tau\sum_{m}(e^{-i\varphi _m }|m\rangle\langle 0m|+e^{i\varphi _m }
|0m\rangle\langle m| ).
%\tau_L\sum_{\alpha}(c_{0\alpha}^{\dagger}d_{\alpha} )+
%d_{\alpha}^{\dagger}c_{0\alpha})\\
%H^{RD}_{{\rm tun}}&=&H^{DR}+H^{RD}
%=\tau\sum_{m}(e^{-i\varphi _m }d_{m}^{\dagger}d_{0m}+e^{i\varphi _m }
%d_{0m}^{\dagger}d_{m} ).
\end{eqnarray}
%Here $\{c^{\dagger}_j,c_j\}$($\{d^{\dagger}_j,d_j\}$) are the pairs of 
%creation and annihilation operators in the leads (interferometer),
Here $\tau_L,\tau$ are the corresponding hopping parameters 
and $0\alpha$ ($0m$) are the nearest sites to the contact points  
$\alpha$ ($m$) between lead-interferometer and ring-dots. 
%In view of further discussion
%we allowed different couplings $\tau_{\alpha}$ for different leads
%(it will be usefull to make the distinction between the lead-dot 
%and the lead-ring coupling constants).   
 
%The effect of the magnetic field is introduced through the 
 $\varphi _m$ is the Peierls phase associated with the pair of sites 
$|0m\rangle$, $|m\rangle$.
Finally $H^D$ is the Hamiltonian of the coupled dots which is 
also written in the Peierls representation. 
It includes the individual Hamiltonian of each dot $H^{D_k}$:
\begin{equation}
H^{D_k}=-eV_k\sum_{i\in QD_k}|i\rangle\langle i|+t_D\sum_{<i,i'>}e^{2\pi i\varphi_{ii'}}
|i\rangle\langle i'| 
\end{equation}
%(spelled explicitely between the braces in the equation below),
and the interdot tunneling term $H_{{\rm tun}}(\tau_{{\rm int}})$,
depending on the coupling constant $\tau_{{\rm int}}$ which is the 
same for each pair of dots $\{k,k+1\}$. 
We point out that the dots embedded in different arms of the ring can be coupled as well,
allowing thus complicated electronic trajectories within the system. 
  The constant term $V_k$ 
from the diagonal part of each $H^{D_k}$ mimics the plunger gate 
voltages used in experiments to tune the dots to resonance, 
$<i,i'>$ denotes the nearest neighbor summation and $t_D$ is 
the hopping integral on dots.

The conductance matrix $g_{\alpha \beta}$ of a mesoscopic system 
at zero temperature coupled to leads is given by the 
Landauer-B\"{u}ttiker formula (see Ref.\cite{JMP} for a 
rigorous derivation)  
\begin{equation}\label{LB}
g_{\alpha \beta}(E_F)=\frac{e^2}{h}T_{\alpha \beta}(E_F)
=4\frac{e^2}{h}\tau_{L}^4\sin^2k|\langle 
\alpha|G_{{\rm eff}}(E_F+i0)|\beta  \rangle |^2,
\qquad \alpha\neq\beta,
\end{equation}
where $T_{\alpha \beta}(E_F)$ is the transmittance, 
$|\alpha\rangle $, $|\beta\rangle $ are sites located on the 
ring or dots that are coupled to the leads, $E_F=2t_L{\rm cos}k$ 
is the Fermi energy of the leads and $t_L$ is the hopping 
integral on leads. 
The main quantity in Eq.\,(\ref{LB}) is the effective resolvent 
of the system in the presence 
of the leads (see Ref.\cite{PRB1} for more details). In our case  
$G_{{\rm eff}}(z)=(H_{{\rm eff}}(z)-z)^{-1}$ where the 
effective Hamiltonian is defined as 
\begin{equation}\label{eff1}
H_{{\rm eff}}(z):=H^I
-\tau_L^2\zeta_1(z)\left (\sum_{\alpha_r}|\alpha_r \rangle \langle \alpha_r |
 +\sum_{\alpha_d}|\alpha_d \rangle \langle \alpha_d |\right )
\end{equation}
and acts in the Hilbert space of the interferometer ${\cal H}_I$ 
only and embodies the influence of the leads at the contact 
points with the ring which we denote $\{\alpha_r \}$ or the 
dots $\{\alpha_d\}$ through the non-hermitian 
terms above (these terms represent the so called leads' self 
energy - see Ref.\cite{Datta}). The notation 
$\zeta_1(z)=(z\mp\sqrt{z^2-4t_L^2})/2$ ($\mp$
shows that $z$ belongs to the upper(lower) half-plane) and we 
choose ${\rm Re}z<2t_L$.
In the sequel we take for simplicity $e=h=t_L=1$.
 
In the previous papers \cite{PRB1,PE} we used simpler effective 
Hamiltonians. In the particular case of a single dot weakly 
coupled to leads (see \cite{PRB1}) Eq.\,(\ref{LB}) 
gives at once the transmittance peaks 
(as a function of the plunger gate voltage $V$) which are 
related to spectral properties of the dot.
Actually the effective Hamiltonian of the dot has resonances with small 
imaginary part located near the eigenvalues of the isolated dot.   
Similarly, if $H_{{\rm eff}}(z)$ describes an array of identical dots 
one can obtain and explain the splitting of the Coulomb peaks 
as a function of the interdot coupling $\tau_{{\rm int }}$ in 
terms of the nearly identical spectra of the dots. 
Moreover, if the interdot Coulomb interaction 
is neglected, the effective Green function can be expressed only in 
terms of one dot 
Green function by a recursive formula. 

Here formula (\ref{LB}) is not of much use because even if  
the transmittance peaks can be obtained from it by inverting 
numerically the finite matrix of the effective resolvent, 
one cannot distinguish between the different paths that an 
electron can follow.
Indeed, due to the ring geometry and to the coupling between the dots 
the transport within the device is very complex.
Besides that, in the experiments the metalic gates defining the dots are 
patterned in the ring arms while the incident electrons from leads enter  
the ring freely.
This means that $\tau_L\sim 1$ thus a discussion in terms of 
the resonances of $H^I$ is useless. 
These drawbacks are only apparent and one can rewrite 
$G_{{\rm eff}}$ in a suitable way to recover the missing details.
The first step is to decompose the Hilbert space of the interferometer 
as ${\cal H}_I={\cal H}_D\oplus {\cal H}_R$ where
${\cal H}_D$(${\cal H}_R$) is the Hilbert space of dots(ring). 
We denote then by $P,Q$ the projectors on these spaces. 
$P,Q$ are nothing else but families of on-site projections 
$|i \rangle\langle i |$  from the coupled dots system and the ring. 
Next, observe that $H^{RD}_{{\rm tun}}$ is a small off-diagonal  
perturbation with respect to 
$H^D-\tau_L^2\zeta_1(z)\sum_{\alpha_d}|\alpha_d \rangle\langle \alpha_d |$
and $H^R-\tau_L^2\zeta_1(z)\sum_{\alpha_r}|\alpha_r \rangle \langle \alpha_r |$ viewed 
as non-hermitian operators in ${\cal H}_D$ and ${\cal H}_R$. This 
allows us to use the Feschbach formula \cite{Fes,RMP} 
which expresses the effective resolvent in the following 
form (see Eq.\,(6.1) from Section VI.B in \cite{RMP})   
\begin{eqnarray}\label{Fes}
G_{{\rm eff}}(z)=G^R_{{\rm eff}}(z)
+(1-G^R_{{\rm eff}}(z)QH_{{\rm eff}}(z)P)\cdot
(H^D_{{\rm eff}}(z)-z)^{-1}\cdot
(1-PH_{{\rm eff}}(z)QG^R_{{\rm eff}}(z) ),
\end{eqnarray}
where we denoted $G^R_{{\rm eff}}(z):=(QH_{{\rm eff}}(z)Q-z)^{-1}$ 
and the new effective Hamiltonian reads 
\begin{equation}\label{HDeff}
%G^R_{{\rm eff}}(z)&:=&(QH_{{\rm eff}}(z)Q-z)^{-1}\\
H^D_{{\rm eff}}(z):=PH_{{\rm eff}}(z)P-PH_{{\rm eff}}Q 
(QH_{{\rm eff}}(z)Q-z)^{-1}QH_{{\rm eff}}(z)P.
\end{equation}
\\
Noticing that in our case $PH_{{\rm eff}}(z)Q=H^{DR}$ one 
obtains by straightforward calculations explicit formulae for 
$G^R_{{\rm eff}}(z)$ and $H^D_{{\rm eff}}(z)$ (we use the notation 
$G_{ij}(z):=\langle i|G(z)|j\rangle $)
\begin{eqnarray}\label{GReff}
G^R_{{\rm eff}}(z)&:=&\left (H^R-\tau_L^2\zeta_1(z)
\sum_{\alpha_r}|\alpha_r \rangle \langle 
\alpha_r |-z\right )^{-1}\\\label{H_eff}
H^D_{{\rm eff}}(z)&:=&H^D-\tau_L^2\zeta_1(z)
\sum_{\alpha_d} |\alpha_d \rangle\langle 
\alpha_d |-\tau^2\sum_{m,m'}e^{-i(\varphi_m-\varphi_{m'})}
G^R_{0m ,0m'}(z) |m \rangle\langle m' |.
\end{eqnarray}
\\
The advantage of using the Feschbach formula is that it provides 
us with two effective resolvents, each one describing 
{\it individually} the pieces that compose the interferometer. 
$G^R_{{\rm eff}}$ describes the truncated ring in the 
presence of the leads while 
$G^D_{{\rm eff}}(z):=(H^D_{{\rm eff}}-z)^{-1}$
is an effective resolvent for the embedded system of dots {\it both} 
in the presence of leads and ring.
We remark that $G^R_{0m,0m'}(z)$ (see Eq.\,(\ref{GReff})) has a 
nonvanishing imaginary part even if $z$ lies on the real axis, 
due to the non-hermitian coupling to the leads. This happens 
because $\zeta_1(E)$ is always complex when $|E|<2t_L$.
By direct computation we express various elements 
of the conductance matrix using Eq.\,(\ref{Fes})
(this time the $E_F$ dependence is omitted as 
well as the subscript '${\rm eff}$' )
\begin{eqnarray}\label{g_1}
g_{\alpha_r, \beta_r}
&=&4\tau_L^4\sin^2k\left |G^R_{\alpha_r,\beta_r} 
+\tau^2e^{i(\varphi_m-\varphi_{n})}G^R_{\alpha_r,0m }G^D_{mn} 
G^R_{0n ,\beta_r} \right |^2=:|t_{\alpha_r,\beta_r}^R+
t_{\alpha_r, \beta_r}^{QD}|^2, \\ \label{g_2}
g_{\alpha_r, \beta_d}&=&4\tau_L^4\tau^2\sin^2k\left |G^R_{\alpha_r,0m }
G^D_{m,\beta_d}  \right |^2, \\ \label{g_3}
% g_{\alpha_d, \beta_r}&=&4\tau_L^4\tau^2\sin^2k
% \left |G^D_{\alpha_d,\beta_r}G^R_{0n,\beta_r }
% \right |^2 \\ \label{g_4}
g_{\alpha_d, \beta_d}&=&4\tau_L^4\sin^2k\left |G^D_{\alpha_d,\beta_d} \right |^2.
\end{eqnarray}
In the above equations the summations over $m$ and $n$ are understood.
The set of formulae (\ref{g_1})-(\ref{g_3}) is the main formal result 
of the paper and the starting point of a detailed discussion of 
the transport processes through the system in terms of the 
spectral properties of the effective Hamiltonian
$H_{{\rm eff}}^D$.  $t_{\alpha_r,\beta_r}^R $ is
the transmission amplitude from lead $\alpha$ to lead $\beta$ via
the truncated ring and $ t_{\alpha_r, \beta_r}^{QD}$ controls the
transport via the arm containing the dot(s). 
In the following we consider  
some particular geometries already used in experiments.

\subsection{One-dot AB interferometer}

 The simplest AB device is realized when there are no leads 
attached to the dots system which in turn is composed of only one dot
(this is the geometry used by Yacoby {\it et al}. \cite{1}). 
Then the term containing the sites $\{\alpha_d \}$ vanishes from 
Eq.\,(\ref{H_eff}) and the transport is completely described by
Eq.\,(\ref{g_1}) that gives the transmittance of the system. 
%It has the form 
%$g_{\alpha_r, \beta_r}=|t_{\alpha_r,\beta_r}^R+
%t_{\alpha_r, \beta_r}^{QD}|^2$ where $t_{\alpha_r,\beta_r}^R $ is 
%the transmission amplitude from lead $\alpha$ to lead $\beta$ via 
%the truncated ring and $ t_{\alpha_r, \beta_r}^{QD}$ controls the 
%transport via the arm containing the dot.

 Let $E_i(V)$ be the $i$-th eigenvalue
of the isolated dot, $\psi_i$ the corresponding eigenfunction and
$P_i:=|\psi _i\rangle\langle \psi _i|$ its associated projection.  
Note that the eigenvalues $E_i(V)$ and their eigenfunctions 
$|\psi _i\rangle$ depend also parametrically on the magnetic field. 
We describe below the resonant transport through $E_i(V)$. 
The idea is to isolate the resonant contribution in the effective 
resolvent.
With the notation $P_i^{\perp }:=1-P_i$, the effective Hamiltonian 
can be written as
\begin{equation}
 H^D_{{\rm eff}}=P_iH^D_{{\rm eff}}P_i
+P_iH^D_{{\rm eff}}P_i^{\perp }+P_i^{\perp }H^D_{{\rm eff}}P_i+
P_i^{\perp }H^D_{{\rm eff}}P_i^{\perp }, 
\end{equation}
and we can apply again the Feschbach lemma
for $(H^D_{{\rm eff}}-z)^{-1}$ having $P_iH^D_{{\rm eff}}P_i^{\perp }
+{\rm h.c}$
as a small perturbation of 
$P_iH^D_{{\rm eff}}P_i+P_i^{\perp }H^D_{{\rm eff}}P_i^{\perp }$.
 Then
with the notations $G_i^{\perp }:=(P_i^{\perp }H^D_{{\rm eff}}P_i^{\perp }-z)^{-1}$
and $D_i(z):=P_iH^D_{{\rm eff}}P_i^{\perp }G_i^{\perp }P_i^{\perp }
H^D_{{\rm eff}}P_i$ the effective resolvent reads
\begin{equation}\label{fesch1}
(H^D_{{\rm eff}}(z)-z)^{-1}=G_i^{\perp }+(1-G_i^{\perp }P_i^{\perp }
H^D_{{\rm eff}}P_i)
(P_iH^D_{{\rm eff}}P_i-D_i-z)^{-1}(1-P_iH^D_{{\rm eff}}P_i^{\perp }G_i^{\perp })
\end{equation}
and the resonant term is clearly
\begin{eqnarray}\label{P_i}
(P_iH^D_{{\rm eff}}(z)P_i-D_i(z)-z)^{-1}=
\frac{|\psi _i\rangle\langle \psi _i|}{E_i(V)-\Delta_i(z)-i\Gamma_i(z)-z}
\end{eqnarray}
where the resonance width $\Gamma_i$ and the shift $\Delta_i$ 
%\begin{eqnarray}\label{selfen}
%\Gamma_i(z):=\tau^2e^{-i(\varphi_m-\varphi_{m'})}\langle \psi _i|m\rangle
%\langle m'|\psi _i\rangle
%G^R_{0m,0m'}(z)-\langle \psi _i|D_i(z)|\psi _i\rangle.
%\end{eqnarray}
are flux-dependent quantities, their expressions being easily 
identified. Notice also that $\langle \psi _i|D_i(z)|\psi _i\rangle$ 
is of order $\tau^4$ thus $\Gamma_i(z)$ is of order $\tau^2$. 
Let now $z\to E_F+i0 $ and suppose that we fix $V$
such that $E_i(V)=E_i-V\simeq E_F+\Delta_i$
($E_i$ being the eigenvalue of the dot in the absence of
the capacitive coupling).  
It is clear that the main contribution in (\ref{fesch1}) comes from 
$(P_iH^D_{{\rm eff}}(z)P_i-D_i(z)-z)^{-1}$ since $G_i^{\perp }$ stays 
bounded and the other terms are of ${\cal O}(\tau^2)$.  
The denominator of $(P_iH^D_{{\rm eff}}(z)P_i-D_i(z)-z)^{-1}$ reduces 
to resonance width $\Gamma_i$ which compensates the multiplicative 
factor $\tau^2$ from the numerator of
$t_{\alpha_r, \beta_r}^{QD}$.  This behavior 
induces a peak in $t_{\alpha_r, \beta_r}^{QD}$ and hence in the total
transmittance across the ring. With these considerations 
we conclude that for weak ring-dot coupling, whenever $V$ comes 
close to $E_F-E_i$ for some $E_i$ the transmittance can be written 
in the form
\begin{equation}\label{phase}
t_{\alpha_r, \beta_r}^{QD}=2i\tau_L^2\tau^2\sin k
G^R_{\alpha_r,0m }G^R_{0n ,\beta_r}\frac{e^{i(\varphi_m-\varphi_n)}
\langle m |\psi_i\rangle\langle\psi_i|
n\rangle }{E_i-V-\Delta_i -E_F-i\Gamma_i}+{\cal O}(\tau^2).
\end{equation}
%are flux-dependent quantities.
% their expressions being easily identified from 
%Eq.(\ref{P_i}).
%$\Delta_i:=\tau^2|\langle \psi _i|m'\rangle |^2
%Re G^R_{0m',0m'} $ and the so-called resonance width
%$\Gamma_i:= \tau^2|\langle \psi _i|m'\rangle |^2
%Im G^R_{0m',0m'}$. We stress that in spite of the fact that 
%$G^R_{0m',0m'}$ is a diagonal element it is 
%{\it not} real because $G^R_{{\rm eff}}$
%contains the non-hermitean term $\zeta_1(z)$ coming from the coupling with the 
%leads. 
Equation (\ref{phase}) is a Breit-Wigner-type formula
and gives the transmittance between the leads via the quantum dot, as 
measured in Refs.\,1-3. A similar formula was obtained by 
Hackenbroich and Weidenm\"uller for a continuous model \cite{HW2,Ha}. 
They supposed that $E_i$ is flux independent, which is true only
at low magnetic fields and small dots. This assumption 
permits an analytical discussion of the flux-dependence of 
$t_{\alpha_r, \beta_r}^{QD}$. As we have said, here we shall not neglect the 
effect of the magnetic field on the dot levels. 
We also point out 
that the 'one resonance' form for the dot transmittance was 
obtained here starting from a many-level description of the dot. 
The rigorous argument for using from the beginning this simplified form 
is that after subtracting the 'resonant' term from the effective 
resolvent the remainder is nonsingular and small.
%The first term in Eq.\,(\ref{g_1}) gives the transmittance in
%the free arm of the ring. Although this part may also exhibit peaks when the 
%Fermi level equals a ring eigenvalue we will not see this in the numerical results 
%given in the next section because
%there is no gate potential applied to the ring and its spectrum does not
%cross the Fermi level.

\subsection{AB interferometer with a coherent double dot}

When $H^D$ describes two coupled dots embedded in different 
arms of the ring connected to two leads (see Fig.\,1) we recover 
the setup of Holleitner {\it et al}. \cite{4}  
In the absence of the lead-dot coupling 
Eqs.\,(\ref{g_2}) and (\ref{g_3}) give no contribution thus 
we are left only with Eq.\,(\ref{g_1}). 
For the simplicity of writing we shall denote
  $G^R_{\alpha _r,0m}:=G^R_{\alpha _r,m}$ and $ \varphi_m-\varphi_{m'}:=\theta_{mm'}$.
Since $G^R_{\alpha\beta}=0$ in this case the conductance has the form:
\begin{eqnarray}\nonumber
g_{\alpha\beta}=4\tau_L^4\tau^4\sin^2k \left |e^{i\theta_{ab}}
G^R_{\alpha a}G^D_{ab}G^R_{b\beta}
+e^{i\theta_{a'b'}}G^R_{\alpha a'}G^D_{a'b'}G^R_{b'\beta}
\right .      \\\label{interf1}
+\left . e^{i\theta_{ab'}}G^R_{\alpha a}G^D_{ab'}G^R_{b'\beta}
+e^{i\theta_{a'b}}G^R_{\alpha a'}G^D_{a'b}G^R_{b\beta}\right |^2 .
\end{eqnarray}
We remark that the terms $G^D_{a'b}$ and $G^D_{ab'}$ connect 
points that belong to different dots. The effective Hamiltonian 
in this case is 
\begin{equation}\label{Dyson1}
H^D_{{\rm eff}}(z)=H^D
-\tau^2\sum_{m,m'}e^{-i\theta_{mm'}}
G^R_{m ,m'}(z) |m \rangle\langle m' |.
\end{equation}
As in the previous section, we are interested in discussing
the resonant transport in terms
of the spectral properties of coupled dots system.
Since the double dot Hamiltonian $H^D$ depends parametrically 
on the capacitive couplings $V_1$, $V_2$
we denote its eigenvalues and eigenfunctions by
$E_i(V_1,V_2)$ and $\psi_i(V_1,V_2)$. 
The main point is that for suitable pairs $\{V_1,V_2\}$
one can bring $E_i(V_1,V_2)$ close to $E_j(V_1,V_2)$ for $j=i+1$.
This is due to the spectral properties of detuned dots.
Let us remind here that the detuning consists in applying an 
additional gate potential to one dot while keeping the other 
gate voltage fixed. In Fig.\,2 we show the spectrum of the
detuned double dot (10$\times$10 sites on each dot) as a
function of the detuning potential $V_{1}$ applied on the first dot,
for a fixed value of $\tau_{{\rm int}}$.
For simplicity the undetuned dot is not capacitively coupled thus
$V_{2}=0$. Obviously, one half of the spectrum shifts linearly
in $V_{1}$.
  The remaining eigenvalues
depend weakly on $V_1$, excepting some points of avoided crossings.
As long as
$\tau_{{\rm int}}\neq 0$ there are no crossings between eigenvalues
(on the contrary, as shown in Fig.\,2 we rather have avoided 
crossings). Moreover, by perturbation theory, near avoided crossings 
the distance between eigenvalues is of order $\tau_{{\rm int}}$.
This behavior of eigenvalues as functions of $V_1$ and $V_2$ is 
due to the fact that, roughly speaking, half of the eigenvalues 
'belong' to QD$_1$, the other half to QD$_2$. As a consequence, when 
$V_1,\,V_2$ are tuned such that both $E_i(V_1,V_2)$ and 
$E_j(V_1,V_2)$ are near and moreover close to the Fermi level 
we expect that {\it both } 
dots will transmit. Clearly one can study
the tunneling through one eigenvalue following the same steps 
as in the analysis of a single dot case. The interesting situation 
is however the one in which the resonant tunneling involves both 
eigenvalues. In the following we show how this appears formally 
at the level of $G^D$. To this end let us introduce
the 2-dimensional projection $P_{ij}:=P_i+P_j$, $P_k$ being the
projection associated to the eigenvalue $E_k(V_1,V_2)$ with $k=i,j$ for
$i$ and $j$ fixed. We shall also use the notation
$P_{ij}^{\perp}:=1- P_{ij}$.
Then $P_{ij}H_{{\rm eff}}^DP_{ij}^{\perp}+{\rm h.c}$ is a perturbation
(of ${\cal O}(\tau^2)$) to $P_{ij}H_{{\rm eff}}^DP_{ij}+
P_{ij}^{\perp}H_{{\rm eff}}^DP_{ij}^{\perp}$ and
by the Feschbach lemma for $G_{{\rm eff}}^D$ one has
\begin{equation}\label{feschbach3}
G_{{\rm eff}}^D=(P_{ij}^{\perp}H_{{\rm eff}}^DP_{ij}^{\perp}-z)^{-1}
+({\tilde H}_{{\rm eff}}^D(z)-z)^{-1}+
{\cal O}(\tau^4),
\end{equation}
with
\begin{equation}\label{Hij}
{\tilde H}_{{\rm eff}}^D(z):=P_{ij}H_{{\rm eff}}^D(z)P_{ij}
-P_{ij}H_{{\rm eff}}^DP_{ij}^{\perp }(P_{ij}^{\perp}
H_{{\rm eff}}^DP_{ij}^{\perp}-z)^{-1}
P_{ij}^{\perp}H_{{\rm eff}}^DP_{ij}=:H^D(z)-D_{ij}(z).
\end{equation}
As in the case of a single dot the first term in (\ref{feschbach3}) 
is small when $z\to E_F+i0$ and the gate voltages are chosen 
such that the resonant condition is fulfilled at least around one 
of the two eigenvalues $E_k(V_1,V_2)$. Then the last step to be
done is to write the Dyson expansion of
$({\tilde H}_{{\rm eff}}^D-z)^{-1}$ taking
$P_i{\tilde H}_{{\rm eff}}^DP_j+{\rm h.c}:=V$ as a perturbation of the
$2\times 2$ diagonal matrix 
$P_i{\tilde H}_{{\rm eff}}^DP_i+P_j{\tilde H}_{{\rm eff}}^DP_j$:
\begin{equation}\label{ds}
{\tilde G}_{{\rm eff}}^D={\tilde G}_{ij,{\rm eff}}^D(z)+{\tilde G}_{{ij,\rm eff}}^D(z)
V{\tilde G}_{ij,{\rm eff}}^D+...,
\end{equation}
where the unperturbed resolvent ${\tilde G}_{ij,{\rm eff}}^D$ has 
the form 
%\begin{equation}\label{redR}
%G_{ij}^D(V_1,V_2,z):=(P_iH_{ij,{\rm eff}}^DP_i-z)^{-1}+(P_jH_{ij,{\rm eff}}^DP_j-z)^{-1}.
%\end{equation}
\begin{equation}\label{redR}
{\tilde G}_{ij,{\rm eff}}^D(z)=
\frac{|\psi _i\rangle\langle \psi _i|}{E_i(V_1,V_2)-\Delta_i(z)-i\Gamma_i(z)-z}+
\frac{|\psi _j\rangle\langle \psi _j|}{E_j(V_1,V_2)-\Delta_j(z)-i\Gamma_j(z)-z}.
\end{equation}
The indices $i,\,j$ were explicitly written for the unperturbed 
operator (we did not introduce another notation).
Thus, we have here two resonances of widths $\Gamma_i,\Gamma_j$ 
(their expressions are complicated but easy to obtain from 
(\ref{Hij})).  
It is clear that as long as the dots are coupled 
$E_i(V_1,V_2)\neq E_j(V_1,V_2)$ thus 
the two resonances come close but do not cross.
Indeed, if for some $V_1,V_2$ the first term in (\ref{redR}) 
behaves like $1/\Gamma_i$ the denominator of the second term is
$\tau_{{\rm int}}+(\Delta_i-\Delta_j)-i\Gamma_j$ thus the 
resonant condition is not strictly achieved. Let us write 
explicitly the second term from the Dyson 
expansion (\ref{ds}). Since $V$ is off-diagonal one is left with:
\begin{equation}\label{ds1}
{\tilde G}_{{ij,\rm eff}}^D(z)V{\tilde G}_{ij,{\rm eff}}^D(z)=
\frac{|\psi _i\rangle \langle \psi _j
 |H_{{\rm eff}}^D+D_{ij}(z)|\psi _i\rangle\langle \psi _j|+{\rm h.c}}
{\left ( 
E_i(V_1,V_2)-\Delta_i(z)- i\Gamma_i(z)-z\right )
\cdot \left (E_j(V_1,V_2)-\Delta_j(z)-i\Gamma_j(z)-z\right ) }.
\end{equation}
Looking at (\ref{Dyson1}) and (\ref{Hij}) one notes that the 
numerator is quadratic in $\tau$ as well as the widths of 
the resonances $\Gamma_i,\Gamma_j$.
Thus the perturbative expansion (\ref{ds}) cannot be used in 
the case of decoupled dots since $E_i$ and $E_j$ can cross, 
the imaginary parts of the resonances are equal 
$\Gamma_i=\Gamma_j=\Gamma$ and 
${\tilde G}_{{ij,\rm eff}}^DV{\tilde G}_{ij,{\rm eff}}^D$
behaves also like  $1/\Gamma$. 
%Thus the   
%the perturbative expansion (\ref{ds}) cannot be used. 
However here we deal with coupled dots and
as long as $\tau_{{\rm int}}>\tau^2$ the Dyson series (\ref{ds}) 
converges and (\ref{interf1}) becomes (omitting the indexes $i,j$ 
and ${\rm eff}$ in ${\tilde G}_{{ij,\rm eff}}^D$)
\begin{eqnarray}\nonumber
g_{\alpha\beta}=4\tau_L^4\sin^2k \left |
\tau^2\left ( e^{i\theta_{ab}}G^R_{\alpha a}\tilde{G}^D_{ab}G^R_{b\beta}
+e^{i\theta_{a'b'}}G^R_{\alpha a'}\tilde{G}^D_{a'b'}G^R_{b'\beta}
\right .\right .      \\\label{paths1}
+\left . \left . e^{i\theta_{ab'}}G^R_{\alpha a}\tilde{G}^D_{ab'}G^R_{b'\beta}
+e^{i\theta_{a'b}}G^R_{\alpha a'}\tilde{G}^D_{a'b}
G^R_{b\beta}+{\cal O}(\tau^2) \right )  \right |^2 .
\end{eqnarray}   
The last formula allows a discussion of the interferometer properties
of the device. The first two terms represent the direct tunneling
through the upper and lower dot, while the terms containing 
$\tilde{G}^D_{ab'}$ and $\tilde{G}^D_{a'b}$ describe paths in 
which the electron tunnels from one dot to
the other before being transmitted in the leads.
At small interdot coupling the cross products 
$\langle a|\psi_k \rangle\langle \psi_k |b'\rangle $, 
$\langle a'|\psi_k \rangle\langle \psi_k |b\rangle $,
$\langle a|\psi_j \rangle\langle \psi_j |b\rangle $ and 
$\langle a'|\psi_i \rangle\langle \psi_i |b'\rangle $ are 
expected to be small so that we can write
(keeping only the first two terms from the right side of Eq.\,(26)) 
\begin{eqnarray}\label{paths2}
g_{\alpha\beta}&=&4\tau_L^4\sin^2k \left |
\tau^2\left ( \frac{e^{i\theta_{ab}}
G^R_{\alpha a}\langle a|\psi_i\rangle\langle \psi_i |b\rangle G^R_{b\beta} }
{E_i(V_1,V_2)-\Delta_i- i\Gamma_i-z} 
+ \frac{e^{i\theta_{a'b'}}G^R_{\alpha a'}
\langle a'|\psi_j\rangle\langle \psi_j |b'\rangle G^R_{b'\beta} }
{E_j(V_1,V_2)-\Delta_j- i\Gamma_j-z}\right )+{\cal R}  \right |^2 , 
%\right .\right .\\\label{paths2}
%&+&\left . \left . {\rm smaller terms} \right )  \right |^2 .
\end{eqnarray}
where ${\cal R}$ collect all the other paths within the interferometer 
that give smaller contributions. Equation (\ref{paths2}) 
will help us to discuss the numerical results from the next section.

One may notice that in the above analysis the spectral properties 
of the truncated ring do not appear in an essential way in 
the problem. This could be anticipated from the beginning
since it is the double dot system that controls the tunneling events.
At the formal level, this fact is revealed only by using Feschbach 
formula.

We mention that our Eqs.\,(\ref{g_1})-(\ref{g_3}) and (\ref{phase}) 
are similar to the ones obtained previously by Hackenbroich and 
Weidenm\"{u}ller \cite{HW1,HW2} by a scattering theory approach, 
in the case of a single dot embedded in a ring connected to two 
leads. Here we gave an alternative calculation in terms 
of the Green functions rather than using the $S$ matrix and we 
generalized the discussion beyond the single-dot case. An advantage of 
our approach is that we do not use the Born series which is formally 
resummed in the scattering approach.  

Let us finally observe that one could not compute the tunneling current 
through the interferometer via rate equation methods used previously 
in \cite{Ulloa} and \cite{Ziegler} for weakly coupled quantum dots. 
These approaches would imply in our problem either the computation of 
the probability distribution$P(N,\alpha)$ characterizing the interferometer 
in the $N$-particle state $\alpha$, either 
a perturbative expansion w.r.t the lead-interferometer tunneling Hamiltonian. 
Since the lead-ring coupling constant is rather big the number of electrons in 
the interferometer is not quantized, thus $P(N,\alpha)$ is not well-defined, 
and the perturbative argument breaks down.

\section{Results and discussion}

We start this section with the most interesting geometry, the one 
used by Holleitner {\it et al}. \cite{4} 
Following their analysis we first look for the charging diagrams of 
a ring with two identical dots connected to two leads.
The dots have 4$\times$5 sites each, while the ring supports 100 sites. 
%We point out that the ratio between the 
%area of the ring and the area of the dot two is much larger that 1
%(in Holleitner's work this ratio is $\sim 100$ ).
We recall (see also \cite{KRMP}) that the charging diagrams 
are plots of the current through a system containing two 
quantum dots as a function of the gate voltages
$V_1$,$V_2$ applied on each dot. 
In Fig.\,3 we present the rhomboids for our system, obtained as
follows: for each fixed value of $V_{2}$, we varied $V_{1}$ 
in the interval shown in the figures and we selected only 
transmittances (i.e. conductances) $T_{12}$ that are larger than 0.4, 
which means that what we obtain is roughly a map for the peak 
positions in the plane $(V_{1},V_{2})$. The magnetic flux is fixed. 
As the interdot coupling increases the diagram changes, due 
to the usual behavior of the transmittance in coupled 
dots \cite{Waugh}: a regular peak is split into two subpeaks, 
separated by a distance which increases
with $\tau_{{\rm int}}$ and saturates at perfect coupling
($\tau_{{\rm int}}=1$). The tunnel split peaks of the interferometer 
transmittance were observed in Ref.\cite{4}
both in vanishing and strong magnetic fields (see Figs.\,4(a) and 
4(b) in the cited reference).    
Figure 4 shows our result for the transmittance of the interferometer  
at uniform capacitive coupling (i.e $V_1=V_2$), 
fixed magnetic flux and different interdot tunneling constants 
(here $\tau_r$ and $\tau$ are also fixed). 
A striking feature is observed in the case of a ring with 
decoupled dots (the dotted line in Fig.\,4): 
the transport is strongly suppressed. This behavior at 
$\tau_{{\rm int}}=0$ was predicted also in Ref. \cite{YK1}. 
It differs from the one encountered in the case of double dots 
connected directly to leads, when two subpeaks merge to a single 
one as $\tau_{{\rm int}}\to 0$.    

Another important aspect of the charging diagram is the drift 
of the peaks near double resonance points, which actually gives 
the honeycomb pattern. We discuss this in
connection with Fig.\,3(b) using the spectral properties of the 
detuned dots emphasized in Section II B. The traces from the range 
$V_2\in (0.11,0.35)$ depend weakly on $V_2$ because the 
corresponding eigenvalues of the embedded double dot 
have this behavior there. A similar behavior is observed
with the traces in the interval $V_1\in (-0.35,-0.11)$ 
where the eigenvalues depend weakly on $V_1$. This behavior 
changes drastically when two traces are approaching
(around point D marked in the figure): they
clearly avoid each other, because the eigenvalues of the 
double dot do not cross ($\tau_{{\rm int}}=0.2$).
The avoided crossing is more difficult to discern at small 
interdot coupling, as in Fig.\,3(a). 
%   Note however that in the grey scale plot presented in Fig.\,2(a) of Ref.\cite{4} 
%  the conductance at the crossing is smaller than the one near the crossing.   
The problem of crossing resonances in double-dot AB interferometers
is discussed in a recent work \cite{W} were it was proved that actually
at real energies such crossings do not exist. This result 
coincides with ours.

In  Ref.\cite{4}, the interferometer properties of the system 
were revealed by the following procedure: for a fixed avoided 
crossing of the charging diagram the current through the 
interferometer was represented as a function of magnetic field. 
We follow the same strategy, by carefully analyzing first what  
happens to the transmittance at such avoided crossing points  
of the charging diagram. As we have mentioned, the two traces 
above regions D and C from Fig.\,3(b) correspond to two eigenvalues 
$E_i(V_1,V_2)$ ($i=1,2$) that depend weakly on $V_2$. 
Similarly, the traces that approach A and D are associated with 
$E_j(V_1,V_2)$. Looking at Eqs.\,(\ref{redR}) and (\ref{paths1}) 
one can notice that as long as $V_{2}$ does not align 
$E_j(V_1,V_2)$ to the Fermi level, 
the only terms that produce peaks in the transmittance
are the ones involving $G^{(i)}$, and this happens each time 
when $E_i(V_1,V_2)\approx E_F$. The main point is that by varying
$V_{2}$ we achieve the resonant condition 
for the term involving $G^{(j)}$, hence both dots will 
transmit.

In Fig.\,5 we show a detail from the charging
diagram in Fig.\,3(b), taken in the neighborhood of almost 
crossing points A and B. In contrast to the usual picture 
with sharp peaks here we observe (Fig.\,5(a)) 
an asymmetric large tail of the peaks, which shows that in this regime
the interferometer acts as a Fano system. This happens 
because one dot (QD$_2$) is always set to a resonance thus 
the corresponding arm of the ring is 'free', providing
the continuum component for the interference.
Formally this is easily understood by looking at 
Eq.\,(\ref{paths2}), because the second term is  
always large enough and interfere with a quantity (the first term) 
that increases as $E_i(V)$ approaches the Fermi level. 
The Fano regime disappears quickly as we tune QD$_2$ 
away from resonance, the picture of separate peaks being recovered 
(Fig.\,5(b)).

In Fig.\,6 a-d the solid lines are plots of the 
transmittance as a function of $V_{1}$ when $V_{2}$ is set 
close to a resonant value.
Remark the sudden drop of the peak after the resonant point 
and the Fano dips. The latter are actually 
located in the avoided crossing region, which explains 
the small transmittance there. Moreover, the asymmetric tail 
changes its orientation as $V_2$ is slightly varied, i.e the Fano parameter sign changes. 
Following Kobayashi {\it et al.} \cite{F1} we shall call this feature the 
electrostatic control of the Fano 
asymmetric line.
%The same phenomena was reported experimentally in Ref.\cite{F1}, 
%when the magnetic field is varied. 
In order to explain this observation we have to look at the two paths that are involved in the 
interference. The first contribution comes from the resonant tunneling through the upper dot
and is given in dashed lines in Fig.\,6 a-d (the plots were obtained by decoupling the the 
lower arm of the ring from the leads). In this case
there is no interference and one gets usual resonant peaks.
 The second contribution is due to the 'background' 
transmittance of the lower arm when $V_2$ is set close to a resonance and upper arm does not transmit.
We illustrate this component of transport in Fig.\, 6e which shows a single peak that appears 
by varying $V_2$ when $V_1$ is far away from resonant values. The points A,B,C,D mark the magnitude 
of the background for four values of $V_2$. Clearly, as $V_1$ approaches the resonant points the 
interference becomes possible and the Fano lines appear. By inspecting each of Figs.\, 6a-d in connection
with Fig.\, 6e one gets a description of the line shape for different pairs of $V_1,V_2$. As long 
as the transmittance values of the two contributions are located on the same side of their 
corresponding peaks the interference is constructive and the Fano line increase up to a maximum which 
coincides with the resonant peak of the upper arm. In contrast, when $V_1,V_2$ are chosen such that 
the transmittance values are located on different sides of the peaks the two path interfere destructively 
and the  Fano line drops to a dip. In particular, for $V_2$ fixed the dips will be located on the same side 
of the peaks, thus the Fano parameter conserve  its sign. 
The appearance of Fano effect
in interferometers with embedded dots was also discussed in a simple 
(exactly solvable) model in Ref.\cite{KK}, without considering the interdot coupling
or emphasizing the electrostatic control of the Fano lineshape.

In the above discussion the magnetic flux was fixed and we have varied $V_2$, emphasizing 
the sensitivity of the Fano interference on this parameter. Fig.\,7 shows that the shape of 
the Fano line can be equally controlled by varying the magnetic flux, while keeping $V_2$ fixed.
Indeed, as $\phi$ increases from 3.00 to 4.50 the asymmetric tail changes its orientation.  
%Figs.\, 7a,b are made at two
%values of $V_2$ that are located on different sides of the background peak. 
This effect originates in the field dependence of the dot levels which 
leads in turn to a shift of the background peak.  
 Indeed from Fig.\,8a one notices at once that the background peak
moves to the left as the magnetic flux is varied. In order to make the connection 
with Fig.\,7  we marked with points the transmittance values corresponding  
to the gate voltage $V_2=0.11$.  As a consequence of the magnetic shift 
the point located at $\phi=3.00$ on the left side of the peak passed on the upper right side 
at $\phi=3.80$ from where it goes down for $\phi=4.50$. The same argument used in the 
discussion of Fig.\,6 explains now the change of the Fano parameter shown in Fig.\,7.

Fig.\,8b shows the $\phi$ - dependence of the resonant eigenvalue of the
{\it isolated} double dot (the line obtained for a vanishing 
ring-dot coupling, i.e $\tau=0$) and of the eigenvalue of the whole interferometer 
(drawn at $\tau=0.3$). The horizontal lines mark the flux values chosen in Fig.\,8a.
As expected a non vanishing $\tau$ leads to a hybridization between the spectra of the truncated
 ring $\sigma (H^R)$ and the coupled dots $\sigma (H^D)$. The double dot eigenvalue acquires a 
quasiperiodic modulation with $\phi$ due to the ring geometry. 

By comparing Figs.\,8a and 8b we observe that 
%up to a shift caused by the real part of the resonance 
the background peak follows the field dependence
of the eigenvalue of the isolated double dot and not the one of the interferometer eigenvalue.
The physical meaning of this behavior is that the resonant transport is controlled by the 
spectral properties of the embedded dots. 
%(recall that in Eq.(\ref{paths2}) the resonant 
%denominators contain precisely the eigenvalues of the isolated double dot).    
 If the interferometer eigenvalue would control the peak position this one should move 
to the right from $\phi=3.00$ to $\phi=3.80$, according to the trajectory 
given for $\tau=0.3$. Clearly this is not the case and, 
up to a shift caused by 
the real part of the resonance the peak obeys the drift of the isolated eigenvalue. 
We stress that this non-trivial effect described above cannot be captured
by a theoretical model that neglects the spectral properties of the dot in magnetic field.
The direction change of the asymmetric Fano tail at the variation 
of magnetic field was experimentally reported by Kobayashi {\it et al}. \cite{F1} in the the 
case of a one-dot interferometer. We believe that the effect we just discussed for the 
two-dots interferometer is similar.      
          
We further investigate the behavior of the Fano peaks as a 
function of the interdot coupling. Figure 9(a) shows that the 
lineshape is very sensitive to this parameter. More interesting 
is the behavior of the interferometer phase along a Fano resonance 
plotted in Fig.\,9(b). For weak coupling (and hence for sharp 
peaks) the phase shows a rapid increase by $2\pi$. 
This feature has some connection with the experimental results  
obtained in a single dot interferometer by Kobayashi {\it et al}.
\cite{F1} They reported an increase of $2\pi$ for the phase of 
the AB oscillations (we present instead the phase of the 
transmittance). In our case the second dot is 
set to a resonance so it acts as a free arm of the ring, 
from where the similarity with the one-dot interferometer. 
By increasing $\tau_{{\rm int }}$ the phase becomes a 
smooth function of $V_1$. 

We now address the problem of AB oscillations. It is clear that they 
are to be observed if both dots are close to resonance, meaning 
that the gate voltages $V_1,V_2$ are suitably tuned 
near some eigenvalues of the double dot. The delicate point is 
that the eigenvalues depend on the magnetic flux through the 
ring so that for different fluxes one needs different resonant 
values for $V_1,V_2$. Otherwise stated, the rhomboids move with 
$\varphi$ (not shown). We found that for small magnetic fields 
the changes are not too drastic and that the AB oscillations can be 
captured by monitoring the Fano dip and plotting the 
transmittance magnitude there as a function of the magnetic flux. 
More precisely, for a given magnetic flux we keep $V_2$ fixed 
and vary $V_1$ in a range that contains only one Fano dip whose  
transmittance is determined (this is simply the lowest value 
in the chosen range). Then we repeat the procedure for other fluxes, 
the results being given in Fig.\,10. 
One can recognize at once the Aharonov-Bohm oscillations.
Their position is slightly shifted due to the phase accumulation 
within the dots (i.e. we express the transmittance as a function 
of the magnetic flux through the ring while the flux encircled 
by the real trajectories is a bit larger).
Notice that the oscillations are in phase at all Fano dips. 
Figure 11 shows that the oscillation amplitude increases as the 
interdot coupling increases.  
 
We have also investigated a single-dot interferometer (the ring has 
the same dimension while the dot is a $8\times 9$ plaquette). 
When the free arm is decoupled (by making some hopping terms zero ) 
we have the usual peaks corresponding to resonant tunneling via the 
dot levels (Fig.\,12(a)). In order to see the Fano features reported 
by Kobayashi {\it et al}. \cite{F1} we restore the coupling to the arm 
and we choose the Fermi level such that the 'background' 
conductance of the arm is around 0.3 (if the Fermi level coincides 
with some eigenvalue of the free arm its conductance approaches unity, 
obscuring thus the contribution of the dot). As expected, the 
symmetric peaks are turned to Fano resonances shown in Fig.\,12(b), 
their correspondence being obvious. One notices that the Fano peaks are either 
wide or very narrow. 
We have checked that this feature remains also valid for other values of the flux and different number 
of sites composing the dot. Remarkably, the Fano parameter 
takes the same sign between succesive peaks. It was suggested recently by Nakanishi 
{\it et al} in Ref.\cite{Nak} that this feature relates to the correlations between the narow and wide peaks.
%We believe that the argument used in the explanation of Fig.\,6 cannot be used as it stands because 
%for the one-dot interferometer the background contribution is not of resonant origin.       

\section{Conclusions}

The main aim of this paper was to present in a unified formalism 
the basic properties of Aharonov-Bohm interferometers with 
 coupled quantum dots. By combining the 
Landauer-B\"uttiker approach and the Feschbach formula we 
studied the transport properties of the interferometer in 
terms of the spectral properties of the embedded dots. Our 
method involves only Green functions and can be viewed 
as an alternative to the scattering theoretical approach.    
In the case of an interferometer with two coupled QD 
(one QD in each arm of the ring) we give a formula (Eq.\,(27)) 
which emphasizes the resonant tunneling  process   
through a given discrete level from the dots (we recall that 
along the paper we have considered many-level dots).

Numerical simulations reproduce the stability charging diagrams 
of two-dot AB interferometer reported in the experiments of 
Holleitner {\it et al}. \cite{4}. 
A careful analysis of the almost crossing points of the diagram 
lead us to several interesting results which are summarized in 
what follows. When the magnetic field is fixed 
and one dot is set to resonance the interferometer transmittance 
shows Fano lineshapes as a function of the gate voltage applied 
to the other dot. 
This
 corroborates with the results of 
%Kubala and K\"onig\cite{KK} obtained in an exactly solvable 
%one-site model and 
shows clearly the coherent 
feature of the transport through the system. We emphasized and explained
the sensitivity of the Fano tail to the gate potential on the second dot.   

As we have said, our model includes the effect of the magnetic field on the 
dot levels. It turned out that this effect explains the change of the 
asymmetric tail as the magnetic flux is varied. It would be of great interest 
to probe experimentally this latter aspect.       
The transmittance 
assigned to the Fano dips shows Aharonov-Bohm oscillations, 
in full agreement with the observation of Ref.\,\cite{4}.  
The influence of the various coupling constants was identified.   
Finally we reproduced the results of Kobayashi {\it et al}.\cite{F1}. 

The analysis of the 4-lead geometry in view of the very recent
results reported in Ref.\cite{4fire} 
is much more complex and requires further investigation.  

\acknowledgments{This work was supported by Grant CNCSIS/2002
and Romanian Programme for Fundamental Research.
V.\,M. acknowledges support from the NATO-TUBITAK and the
Romanian Ministry of Education and Research
under CERES contract . B.\,T. acknowledges the support of
TUBITAK, NATO-SfP, MSB-KOBRA, and TUBA. The authors are 
very grateful to Ulrich Wulf for many valuable discussions.}

\newpage

\begin{figure}
\caption{Schematic picture of a two-dots Aharonov-Bohm interferometer. The thick solid line 
represents the truncated ring (R). The dashed countour surrounds the interferometer (I). 
$\alpha$, $\beta$ are the sites where the leads are connected to the interferometer and
$a,a',b,b'$ are the contact points between ring and dots.}
\end{figure}

\begin{figure}
\caption{Avoided crossings in the spectrum of a $20\times 10$ 
double dot as a function of the detuning potential $V_1$ 
applied on QD$_1$ ($\tau =0.4$, $\Phi =0.15$, $\tau_{{\rm int}}=0.1$). Here $\Phi$
 is the magnetic flux through one cell, in flux quanta.}
\label{figure4}
\end{figure}

\begin{figure}
\caption{Charging diagrams for the double dot Aharonov-Bohm 
interferometer ($\tau_L=1$, $\phi=3$, $\tau =0.3$). $\varphi$ 
is the magnetic flux through the ring, in flux quanta. The traces 
represent transmittances bigger than $0.4$. (a) $\tau_{{\rm int}}=0.1$,
(b) $\tau_{{\rm int}}=0.2$,
(c) $\tau_{{\rm int}}=0.5$. The Fermi level is set to 0.}
\label{figure3}
\end{figure}

\begin{figure}
\caption{The effects of the interdot coupling $\tau_{{\rm int}}$ 
on the electronic transmittance of a double dot Aharonov-Bohm 
interferometer at fixed magnetic flux $\phi=3$. The same
gate potential $V$ is applied on each dot $\tau_L=1$, $\tau=0.5$. 
Full line - $\tau_{{\rm int}}=1$, 
long dashed line - $\tau_{{\rm int}}=0.5$,
dashed line - $\tau_{{\rm int}}=0.2$,
dotted line - $\tau_{{\rm int}}=0$ (transport is strongly suppressed).}
\label{figure2}
\end{figure}

\begin{figure}
\caption{The structure of the transmittance peaks from Fig.\,3(b) 
around the points of double resonance (a). Away from this point 
one has distinct sharp peaks that turn into Fano peaks at 
the avoided crossing points (b).}
\label{figure5}
\end{figure}

\begin{figure}
\caption{ a)-d) Solid lines: Fano line shapes as a function of $V_1$ for several 
values of $V_2$. The Fano tail changes its orientation by passing through a symmetric maxima.
Dashed lines: The resonant transport through the upper arm of the ring when the lower arm is decoupled
from leads. Remark the correspondence between the usual peaks and the Fano maxima. 
 (a) $V_2=-0.1175$, (b) $V_2=-0.1150$, (c) $V_2=-0.1125$, (d) $V_2=-0.1100$ . e) A resonant peak as a function of 
$V_2$. The gate potential on $QD1$ was set to $V_1=-0.2$. The points A,B,C,D correspond to the 
values of $V_2$ chosen in Figs.\,a)-d). All plots are made for $\tau=0.3$, $\tau_L=1$, $\tau_{{\rm int}}=0.2$,
$\phi=3$.}
\label{figure6}
\end{figure}

\begin{figure}
\caption{(Color online). Magnetic control of the Fano interference. As the magnetic field is 
varied the Fano parameter changes sign. 
%The two gate voltages on the second dot were chosen on 
%different sides of the peak. 
 $\tau=0.3$, $\tau_L=1$, $\tau_{{\rm int}}=0.2$.}
\label{figure7}
\end{figure}

\begin{figure}
\caption{ (Color online). a) The background peak moves with the magnetic flux;
the gate potential on $QD1$ was set to $V_1=-0.2$ . b) The eigenvalue of the 
decoupled double dot ($\tau=0.0$) has a positive slope w.r.t the magnetic flux. The interferometer 
eigenvalue ($\tau=0.3$) is additionally modulated by the hybridization between the truncated ring
and the double dot.}
\label{figure8}
\end{figure}

\begin{figure}
\caption{The sharpness of the Fano resonances (a) and the phase of the
transmittance through the interferometer (b), as a function of the interdot coupling: 
full line - $\tau_{{\rm int}}=0.05$,
dashed line - $\tau_{{\rm int}}=0.15$, 
dotted line - $\tau_{{\rm int}}=0.3$.
At weak coupling the phase increases rapidly by $2\pi$ while for stronger 
coupling it increases smoothly by $2\pi$ along a Fano resonance. 
The parameters used are $V_2=-0.110$, $ \phi=3$, $\tau=0.3$,
$\tau_L=1$.}
\label{figure9}

\end{figure}
\begin{figure}
\caption{The in-phase Aharonov-Bohm oscillations of the transmittance 
assigned to the Fano dips from the region A, B, C, and D in 
Fig.\,3(b).} 
\label{figure10}
\end{figure}

\begin{figure}
\caption{Aharonov-Bohm oscillations in the region D of the 
charging diagram at different interdot couplings: full line - 
$\tau_{{\rm int}}=0.25$, dashed line - $\tau_{{\rm int}}=0.2$, 
dotted line - $\tau_{{\rm int}}=0.15$. Other parameters are
$t_L=1$, $\tau=0.3$, $E_F=0$. }
\label{figure11}
\end{figure}
  
\begin{figure}
\caption{Transmittance through a single dot 
interferometer ($\tau_L=1$, $\tau=0.35$, $\phi=5$, $E_F=-0.5$ ). 
The dot has $9\times8$ sites and the ring contains 100 sites. 
(a) Usual peaks arising from the resonant tunneling via the discrete 
levels of the dot (the free arm of the ring is decoupled).
(b) The Fano regime: the free arm conducts and interferes with 
the path along the QD. The peaks turn to Fano lineshapes.}  
\label{figure12}
\end{figure}

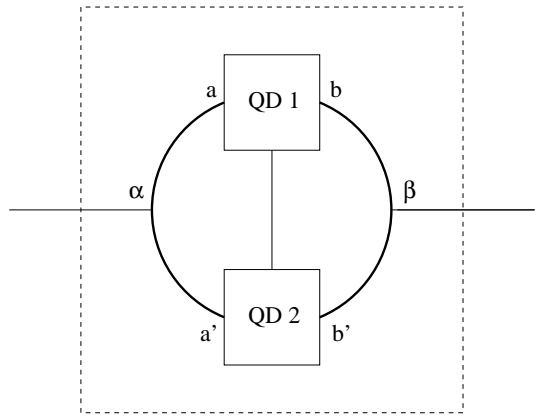
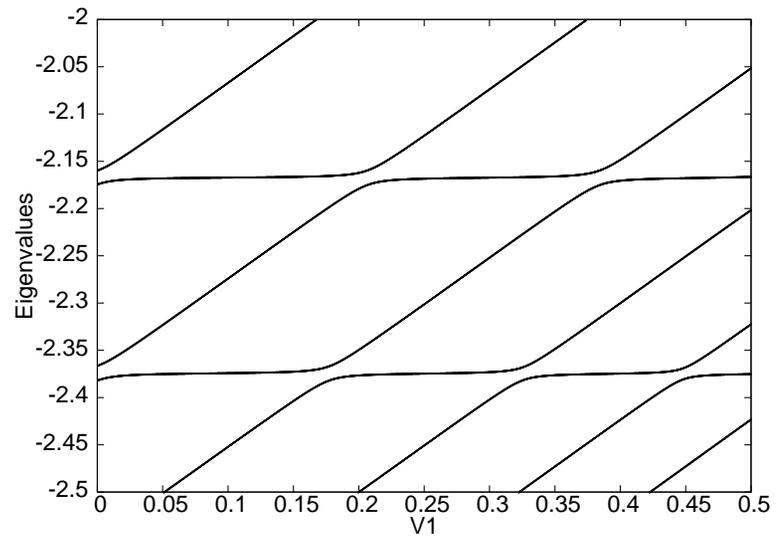
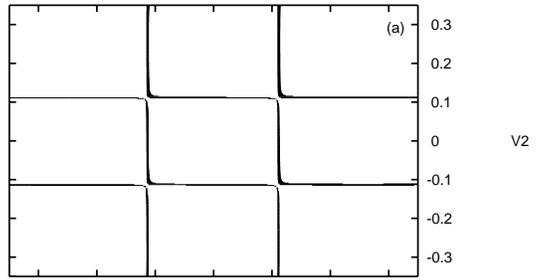
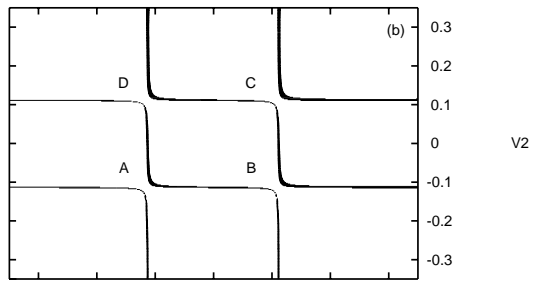
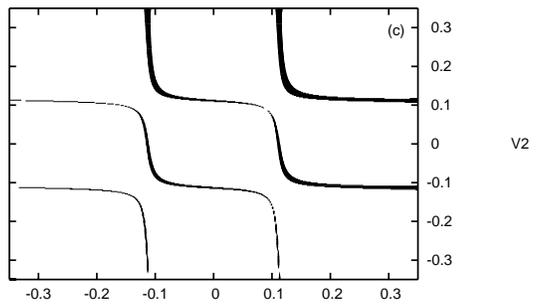
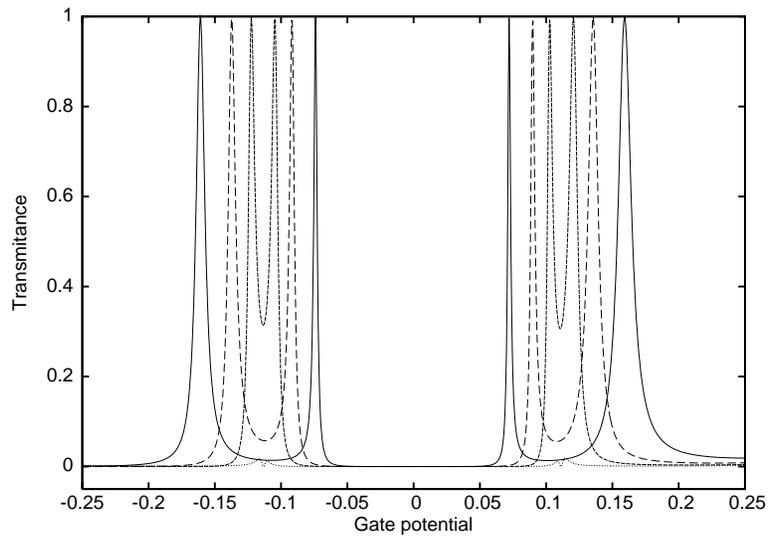
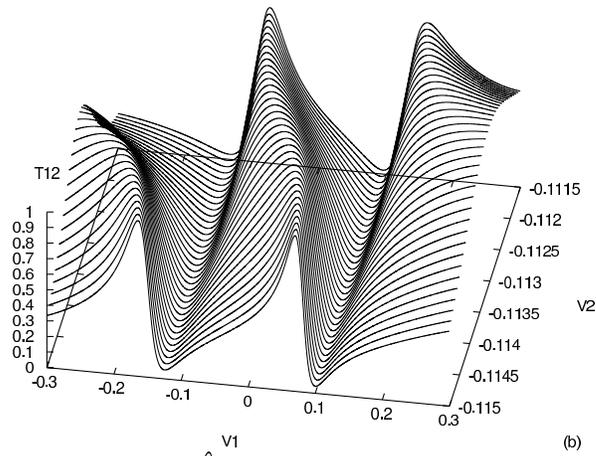
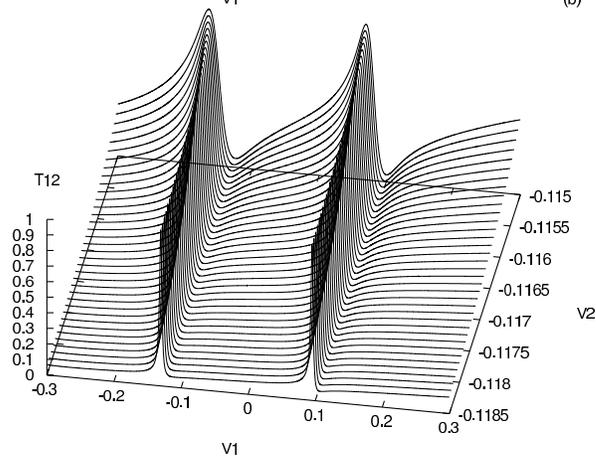
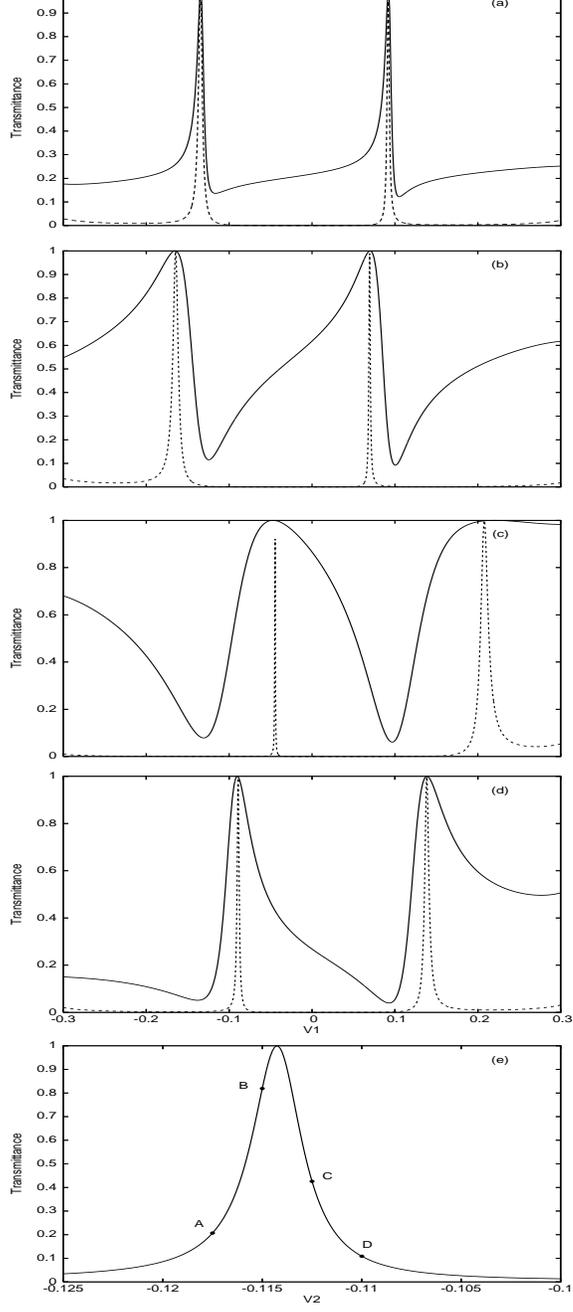
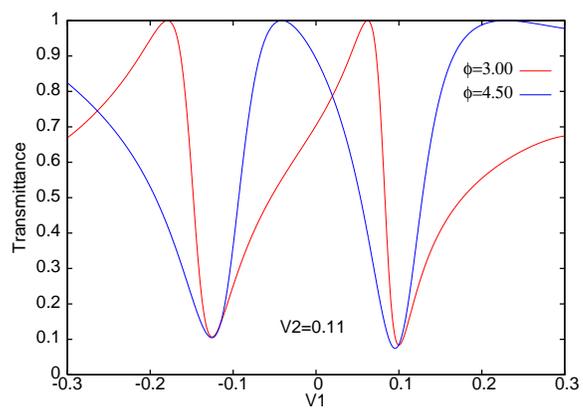
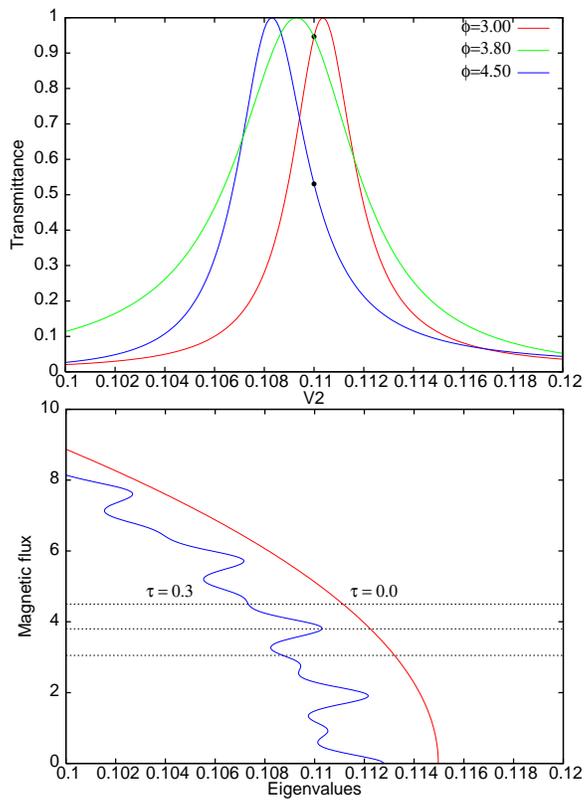
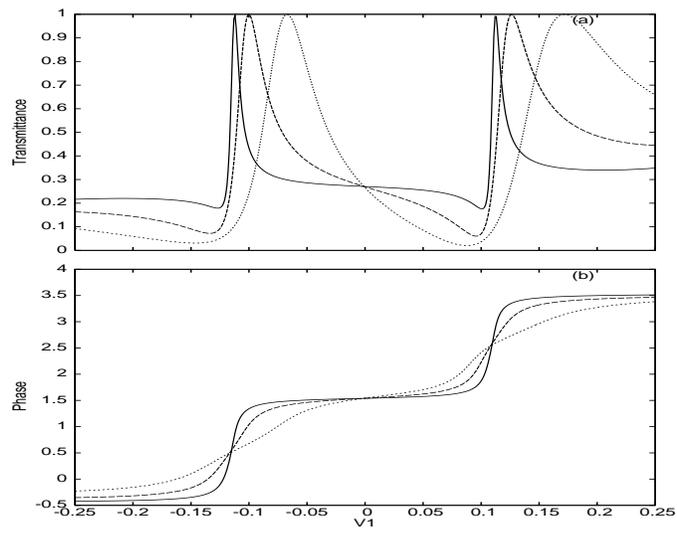
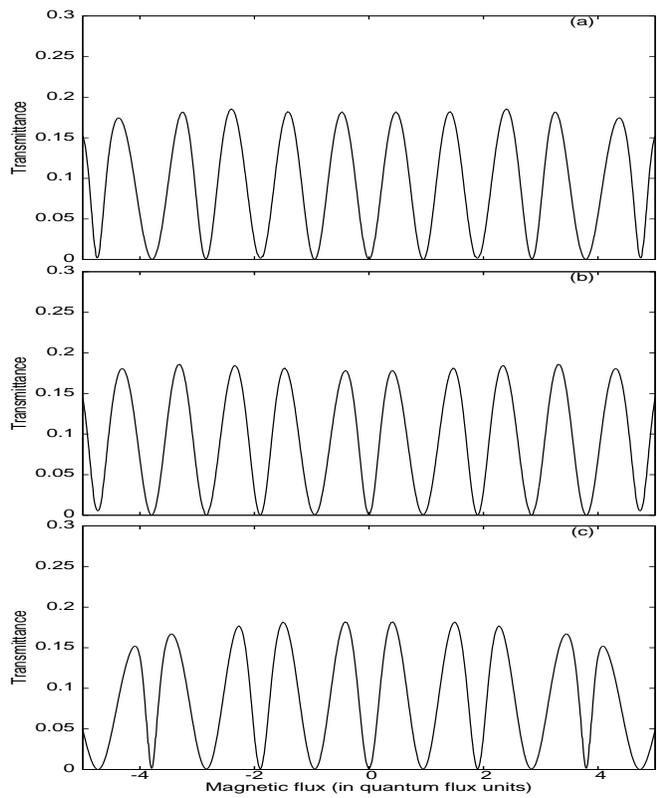
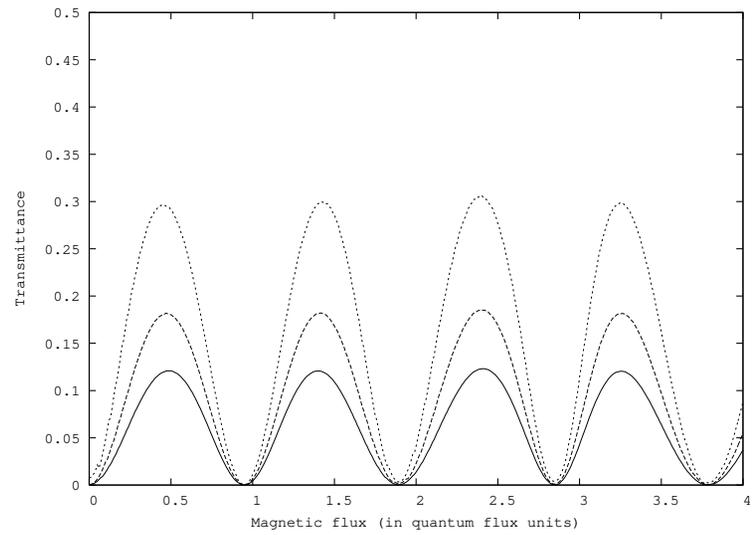
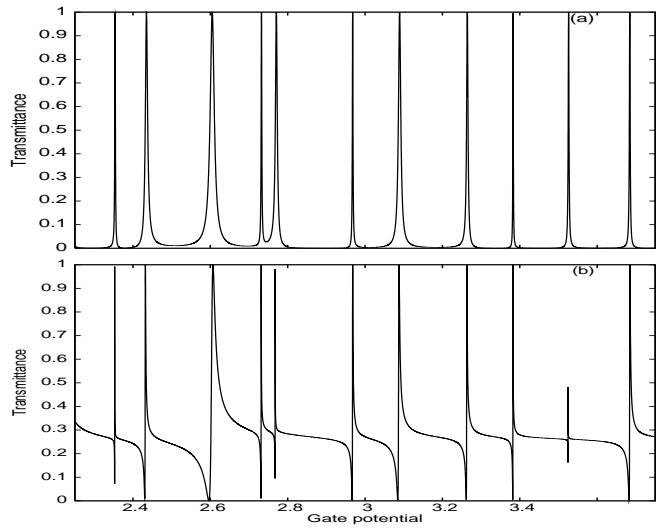
\end{document}